\documentclass{PoS}
\usepackage{soul}

\def\beqar{\begin{eqnarray}} 
\def\eeqar{\end{eqnarray}}

\def\pfrac#1#2{\left( \frac{#1}{#2} \right)} 
\def\avg#1{\langle #1 \rangle}

\title{Rapid variability at very high energies in Mrk 501}

\ShortTitle{Rapid VHE variability in Mrk 501}

\author{\speaker{Nachiketa Chakraborty}\,$^a$\,\thanks{Fellow of the Alexander von Humboldt Foundation} , Gabriele Cologna\,$^b$\thanks{Member of the International Max
    Planck Research School for Astronomy and Cosmic Physics at the University
    of Heidelberg (IMPRS-HD) and the Heidelberg Graduate School of Fundamental
    Physics (HGSFP).}, Max Anton Kastendieck\,$^c$, Frank Rieger\,$^d$, Carlo Romoli\,$^e$
  , Stefan J. Wagner\,$^b$,Agnieszka Jacholkowska\,$^f$, Andrew Taylor\,$^d$ for the H.E.S.S. Collaboration \\
        \llap{$^a$} Max-Planck-Institut f\"ur Kernphysik, Saupfercheckweg 1, 69117 Heidelberg, Germany \\
        \llap{$^b$} Landessternwarte, Universit\"at Heidelberg, K\"onigstuhl, 69117 Heidelberg, Germany \\
         \llap{$^c$} Universit\"at Hamburg, Institut f\"ur Experimentalphysik, Luruper Chaussee 149, D 22761 Hamburg, Germany \\
        \llap{$^d$} ITA Universit\"at Heidelberg and Max-Planck-Institut f\"ur Kernphysik, Heidelberg, Germany \\
        \llap{$^e$} Dublin Institute for Advanced Studies, 31 Fitzwilliam Place, Dublin 2, Ireland\\
        \llap{$^f$} LPNHE, Universit\'e Pierre et Marie Curie Paris 6, Universit\'e Denis Diderot Paris 7, CNRS/IN2P3, 4 Place Jussieu, F-75252, Paris Cedex 5, France\\
     	Email: \email{cnachi@mpi-hd.mpg.de}}

\abstract{A major flaring state of the BL Lac object Mrk 501 was observed
  by the High Energy Stereoscopic System (H.E.S.S.) in
    June, 2014. Flux
  levels higher than one Crab unit were recorded and rapid variability at very
  high energies ($\sim$2-20\,TeV) was revealed. The high statistics afforded
  by the flares allowed us to probe the presence of minutes timescale
  variability and study its statistical characteristics
  exclusively at TeV energies owing to the high energy threshold of
  approximately 2\,TeV. Doubling times of a few minutes are estimated for
  fluxes greater than 2 TeV. Statistical tests on the light
  curves show interesting temporal structure in the variations including
  deviations from a normal flux distribution similar to those found in the PKS
  2155-304 flare of July 2006, at nearly an order of magnitude higher
  threshold energy. Rapid variations at such high energies put strong
  constraints on the physical mechanisms in the blazar jet.} 

\FullConference{The 34th International Cosmic Ray Conference,\\
		30 July- 6 August, 2015\\
		The Hague, The Netherlands}

\begin{document}
\section{Introduction}

Markarian\,501 (Mrk\,501) at $z=0.034$ is a well studied high-frequency peaked
BL Lacertae (HBL) object. Historically, it 
has shown highly variable emission in wavelengths ranging from radio to very
high energy (VHE, E\,>\,100\,GeV) gamma-rays. It was 
first detected above 300\,GeV by
the Whipple Observatory in 1996 \cite{1996ApJ...456L..83Q}. Since then it has
been observed several times at TeV energies 
\cite[etc.]{1999A&A...349...29A, 2009arXiv0912.3772H,2007ApJ...669..862A}. It has shown VHE variability  
down to minute timescales between 0.15-10\,TeV as reported by MAGIC
in \cite{2007ApJ...669..862A}. TeV observations coupled with X-ray observations
usually put strong constraints on the source magnetic field and Doppler factors
within the context of a homogenous synchrotron self-Compton model
\cite{1998ApJ...509..608T}. Observations by the High Energy
Stereoscopic System (H.E.S.S.) have been carried out in four periods between
2004 and 2014 \cite{2015arXiv1310.1201L}. In this proceeding, the focus will
be on the 2014 observations, which includes the strongest flare ever detected
by H.E.S.S. Afforded by the unique data set, the VHE variability is
investigated. It is at energy scales higher than reported before and the aim
is to improve and extend constraints on the source emission mechanisms. Also,
comparisons with the 2006 flare of PKS 2155-304
 \cite{2010A&A...520A..83H} are instructive in looking
for general features in the temporal structure of the emission and therefore
for potential principles underlying the physics of jet emission.

\section{Rapid flux variations at very high energies}
The 2014 H.E.S.S. observations of Mrk\,501 at large Zenith
  angles (>60$^\circ$) were performed as a target of opportunity
following fluxes over 1 Crab unit reported by the FACT
  collaboration. Nightly fluxes above 2\,TeV
ranging from $\sim$3 to 40$\times$10$^{-12}$\,cm$^{-2}$s$^{-1}$ were recorded between the nights of June
19-25. On the night of June 23-24 a large flare was detected 
comprising the highest fluxes
of Mrk 501 recorded by H.E.S.S. This study focuses on the temporal
characteristics of the 2014 data and explores physical constraints that
may be derived with time series analysis studies. 
More information about the spectral analysis of the whole dataset as
  well as the multiwavelength context can be found in \cite{2015arXiv1310.1201L}. 

The observations during this period comprise of data from all five
H.E.S.S. telescopes CT1 to CT5. In order to
  have an homogeneous dataset and eliminate variance due to two telescope
  types, only information from CT1-4 has been extracted from the data. The livetime is $\sim$7.7 hours (6.8\, after acceptance correction). The mean Zenith
angle is 63.7$^{\circ}$. The analysis was performed
with the Model analysis method \cite{2009APh....32..231D} with $Loose$ cuts. The background estimation was
made using the $Reflected\ Region\ Background$ method
\cite{2007A&A...466.1219B}.  
As a result of the high Zenith angles, the energy threshold
is $\gtrsim$\, 2\,TeV. Therefore, the
striking feature of these observations is that the light
curve obtained is exclusively at TeV energies. Consequently, the
variability detected is also exclusively at TeV energies, unlike
previous studies where the variations were dominated by fluxes at
energies of few hundreds of GeVs. 

\begin{figure}[h!]
  \centering
 
  \includegraphics[width=1.0\linewidth]{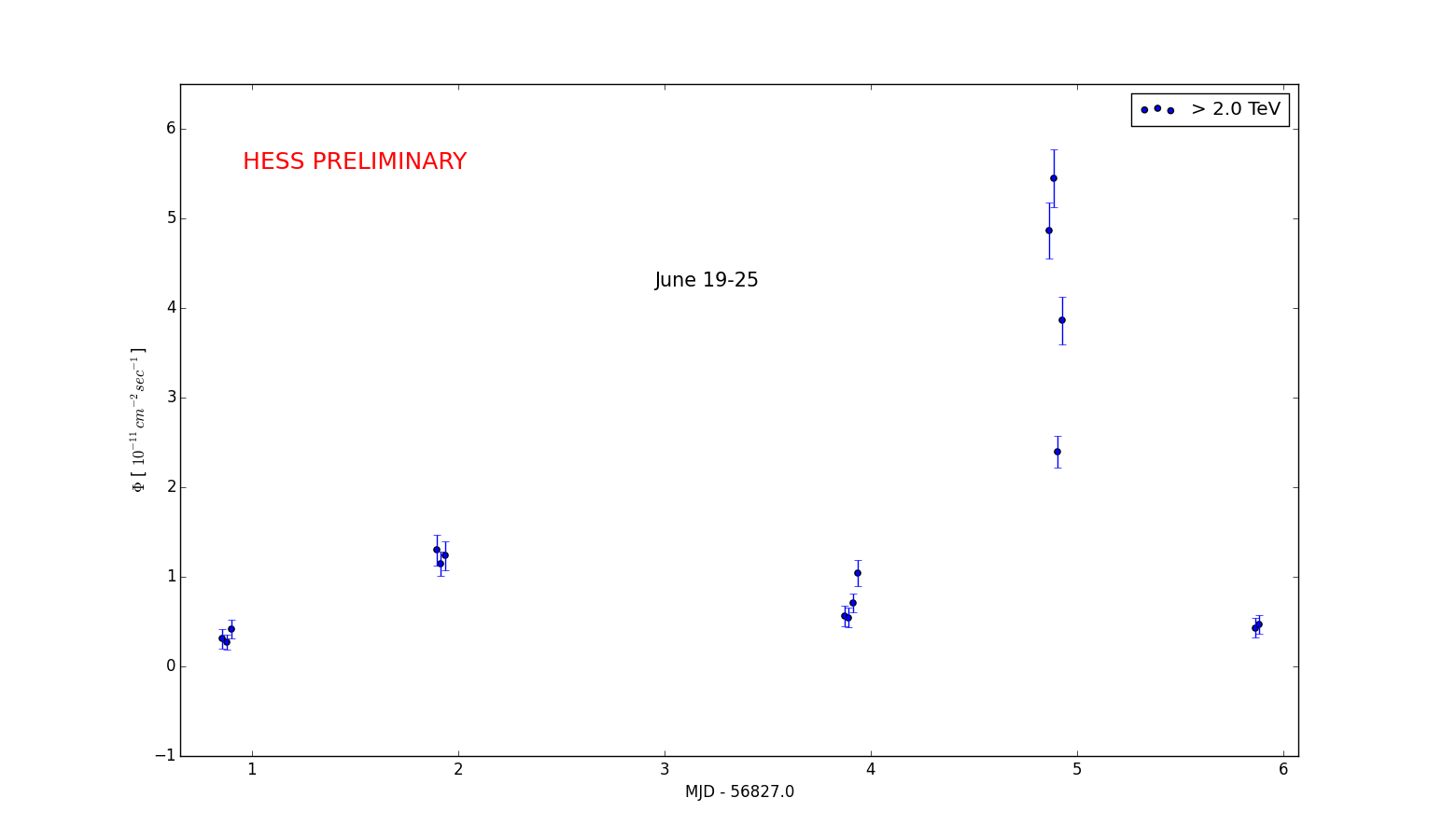}
  \caption{The figure shows the variations in the flux for Mrk~501 over the course of the few days of activity in June 2014. The 
  light curves are binned runwise. The fluxes are above the analyses threshold of 
   $\approx$ 2 TeV. The flux points in the figure are shown with 1-$\sigma$ error bars.}
  \label{fig:fullflaremrk5012014}
\end{figure}

\begin{figure}[h!]
  \centering
  \includegraphics[width=1.1\linewidth]{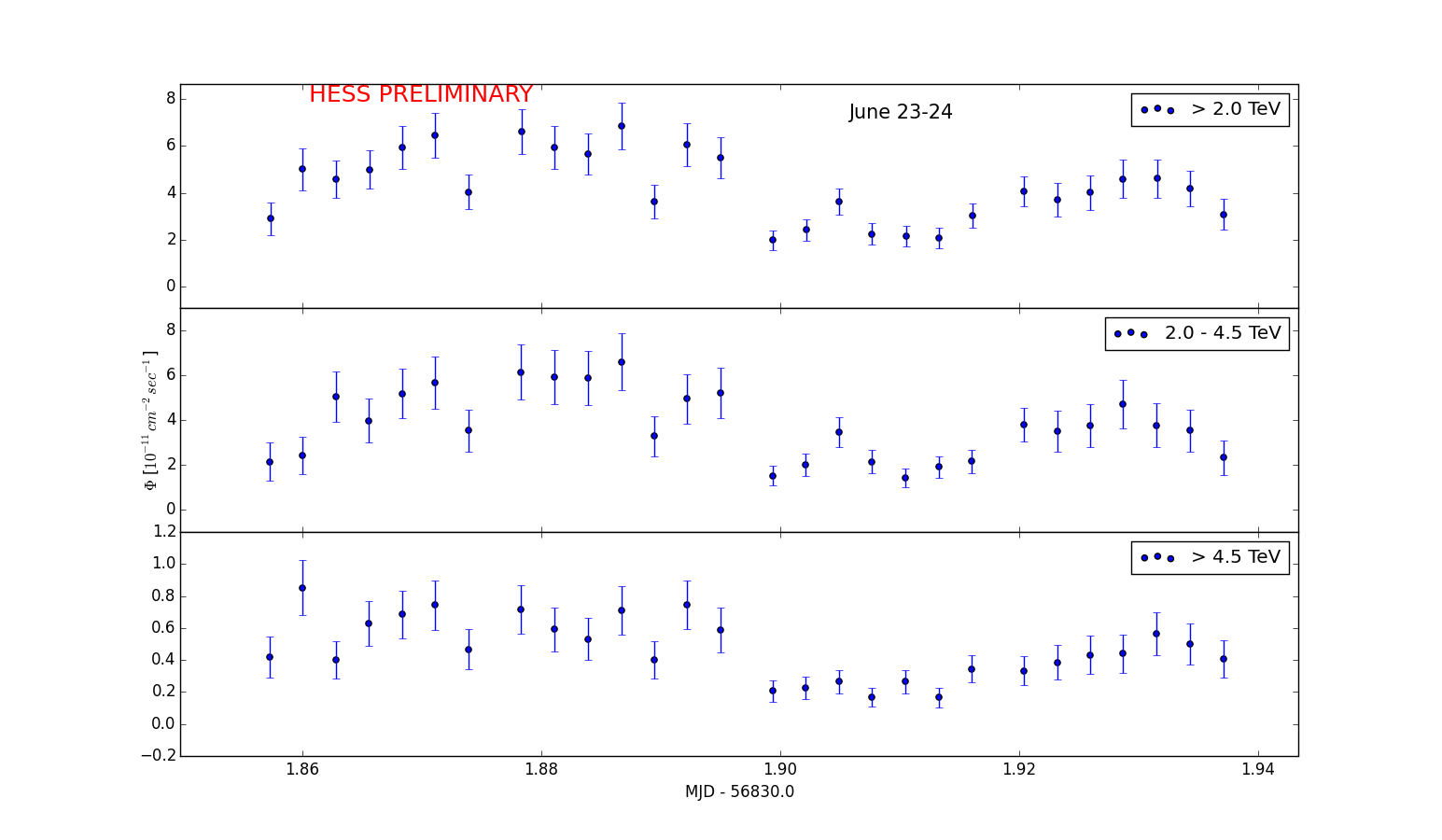}
   \caption{The figure shows a zoom into the peak of the flare on June 23-24. As in the previous figure, the panels are $\phi$(> 2 TeV), 
   $\phi$(2-4.5 TeV) and $\phi$(>4.5 TeV). It is evident that there is variation on a few minute timescales at every energy range 
   including the one above 4.5 TeV. The 1-$\sigma$ error bars are shown. }
  \label{fig:peakflaremrk5012014}
\end{figure}

Variations at TeV energies are found down to a few minutes (<\,10 minutes),
as shown in Fig.~\ref{fig:fullflaremrk5012014}. The
lightcurve covers the days from MJD
56828 to 56833 and shows a strong
flaring event during MJD 56831-56832.
A zoom into the peak of the flare
state is shown in 
Fig.~\ref{fig:peakflaremrk5012014}.
 Using the median energy of the photons above 2\,TeV the lightcurve is divided into 2 bands: $2.0-4.5$\,TeV and $E >
4.5$\,TeV. As shown in Fig.~\ref{fig:fullflaremrk5012014} and more sharply in Fig.~\ref{fig:peakflaremrk5012014}, short
timescale variations are seen at energies $>2$\,TeV as well as
$>4.5$\,TeV. The fractional variability $F_{\rm var}$ is used to quantify timescales 
following \cite{2003MNRAS.345.1271V}. It is the excess variance scaled down 
by the mean flux as, 
\beqar
\label{eq:fvar}
F_{\rm var} = \sqrt{ \pfrac{\phi^{2} - \avg{\sigma_{\rm err}^{2}} }{\phi_{\rm mean}^{2}} }
\eeqar
The values are obtained for
each of the bands shown in the figures. For $E>2$\,TeV, $F_{\rm var} \sim
1.1\pm0.3\times10^{-1}$. 
Such short timescale VHE variability has also been observed by H.E.S.S. in another HBL source, PKS\,2155-304 ($z=0.116$), whose
temporal structure has been studied in detail \cite{
  2010A&A...520A..83H, 2007ApJ...664L..71A}. 
 The flaring state of PKS
  2155-304 in July 2006 had higher VHE fluxes than those of
  Mrk\,501 in June 2014; this included two nights, MJD 53944 and MJD 53946
  where the flux levels were $\sim$\,9 and $\sim$\,12 times that of the Crab
  Nebula at these energies. Therefore, constraining the temporal
  structure for the Mrk\,501 flares and thus the VHE emission mechanisms
  appears more challenging. On the other hand, the spectral characteristics imply that the flare in PKS\,2155-304
was dominated by fluxes at energies lower than 1\,TeV.
In the following, a preliminary
investigation of the variability characteristics of Mrk\,501 
will be made, together with the comparison,
when appropriate, with the results
for PKS\,2155-304 reported in \cite{2010A&A...520A..83H} in order to
understand if the emission mechanisms are similar and
possibly extend over a broad range of energies. Some techniques used for the temporal studies of PKS\,2155-304 are used
here. Flux doubling times defined as in \cite{1999ApJ...527..719Z} are used as
an estimator for the variability timescales. It is defined as 
\beqar
\label{eq:fluxdoubtime}
T_{\rm 2}^{j,k} = | \bar{\phi}\ \Delta T / \Delta \phi |
\eeqar
where $\Delta\phi = \phi_{j}-\phi_{k}$ and $\Delta T = T_{j}-T_{k}$ are the
flux and time differences between the {\it j}th
and {\it k}th data points,
respectively, and $\bar{\phi} = \frac{\phi_{j} + \phi_{k}}{2}$ is the
corresponding mean flux. The entire data set from June 19-25th is used for
this analysis. The time binning is selected ensuring a minimum significance ($\gtrsim
 3\sigma$) per bin during the flare.

The minimum variability timescale computed by this method is $T_{\rm min} < $ 10 minutes. The minimum of the values of 
time difference between all pairs $T_{2,\rm min}$, is computed to be $\approx 6.5\pm2.9$ minutes. The mean of the 5 smallest values of these 
pairwise time differences is $\tilde{T}_{2,\rm min} \approx 6.6\pm2.8$ minutes. The pairs chosen have a relative error $\leq 50\%$ obtained by propagating the error in the fluxes. 
This gives a rough estimate of the variability timescales, even though it is understood that this is not fully robust and sensitive 
to the time binning. In this case the bin size used was 4 mins. So far, variations at such timescales observed in Mrk\,501
were dominated by energies
below the threshold of this observation
($\sim$2\,TeV). Figure~\ref{fig:peakflaremrk5012014} shows a zoom-in to the
peak of the flare. This clearly shows variability at a few minute timescales
for both fluxes above 2 and 4.5 TeV.

\section{Underlying TeV flux distribution} 
\begin{figure}[h!]
  \centering
\includegraphics[width=0.45\linewidth]{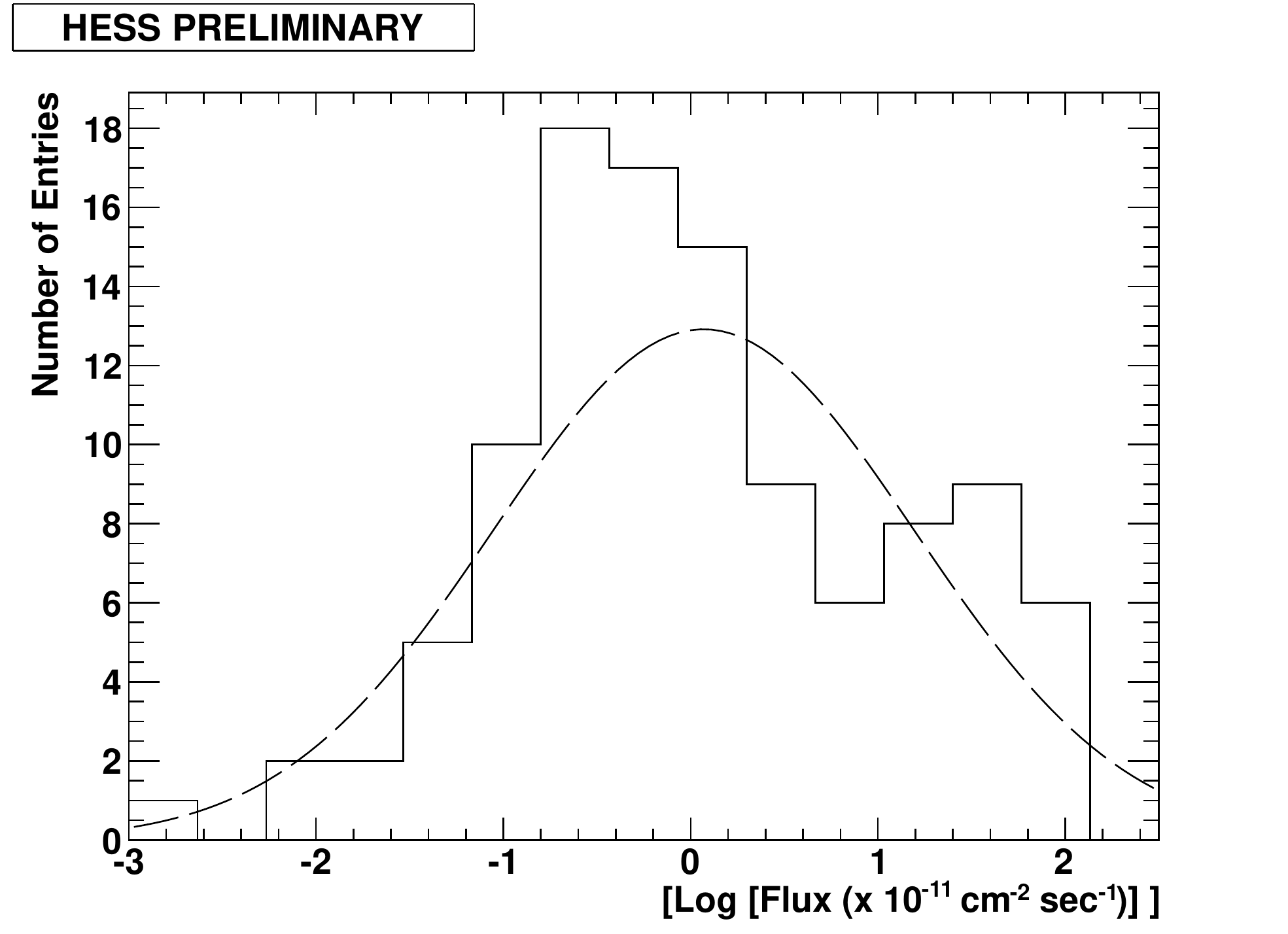}
  \includegraphics[width=0.45\linewidth]{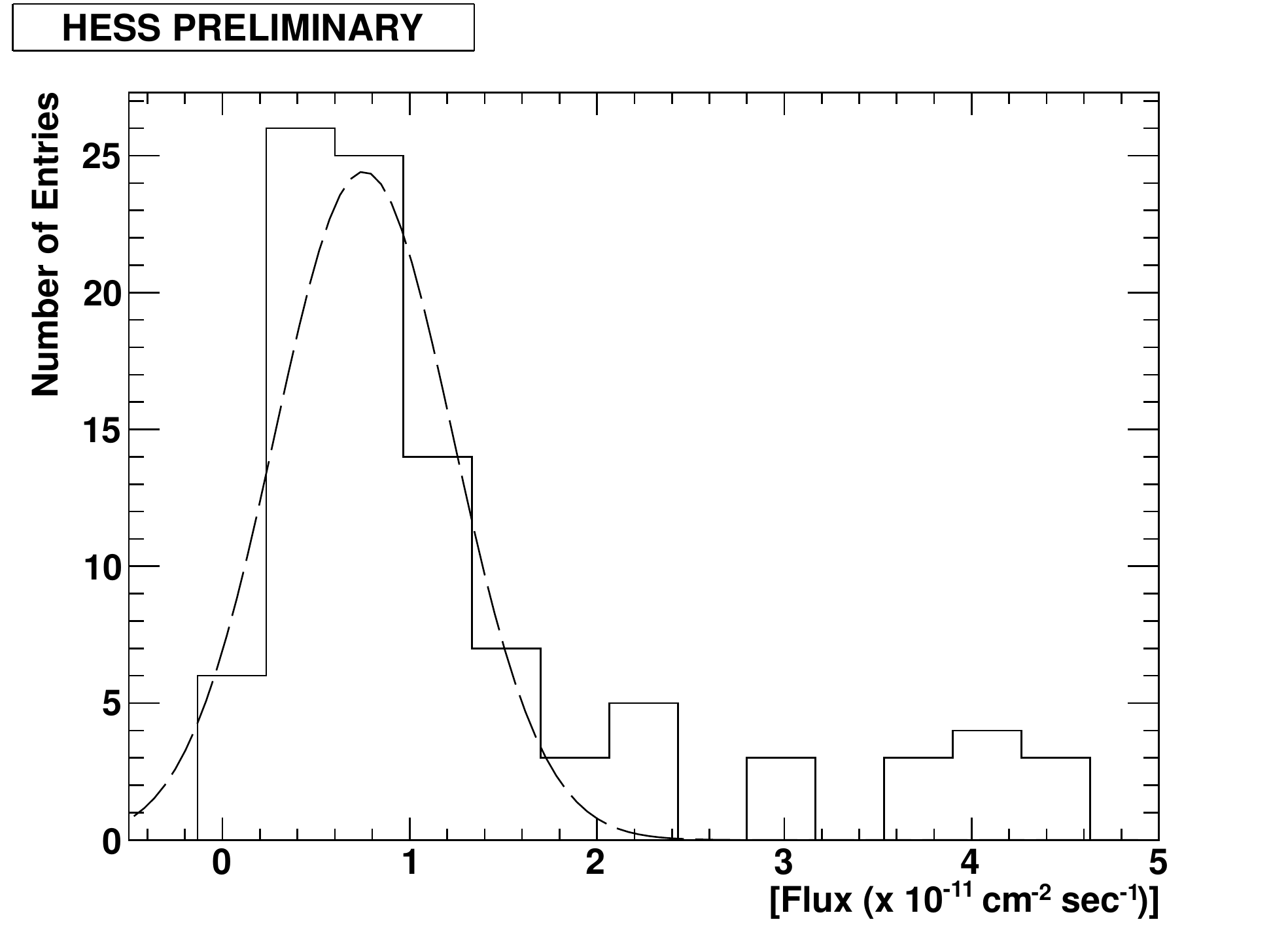}
  \caption{The figure shows the distribution of TeV fluxes from the entire duration of the flare in Mrk~501. To the left is the 
  distribution of the fluxes and to the right the logarithm of the fluxes. There is a statistical preference towards a log-normal 
  distribution as suggested by the reduced chi-squared values. This hints to a lognormal behaviour at TeV energies.}
  \label{fig:pdfmrk5012014}
\end{figure}

The underlying flux probability distribution function (PDF) gives hints as to the emission mechanisms. The PDF is represented simply by the histogram of the fluxes. If the PDF is lognormal, this suggests a 
multiplicative emission process as opposed to an additive process. In principle, this would be a characteristic of a "cascade-like'' 
process. For PKS 2155-304 strong evidence for log-normality was found in the July 2006 flare data and in general in multiwavelength data \cite{2015arXiv1310.1200L}, revealing a clear preference 
for a normal distribution of the logarithm of fluxes rather than the fluxes themselves. This could possibly point to very interesting 
physical implications \cite{2008bves.confE..66S, 2010A&A...520A..23R}. Thus, the distribution of TeV fluxes shown 
in figure~\ref{fig:fullflaremrk5012014} has been tested for the same effect in Mrk~501. It is found that even in this case, fluxes tend 
to prefer a log-normal distribution as shown in figure~\ref{fig:pdfmrk5012014}. This is based on a 
chisquare fit of both the distribution of fluxes as well as the logarithm of fluxes using the Gaussian function of form,  $\frac{A}{\sqrt{2\ \pi\ \sigma^{2}}}\exp(-\ (x-\mu)^{2} / 2\ \sigma^{2})$. There are 15 flux bins in each case with 3 bins per linear (or logarithmic) unit interval. The values of the reduced chi-square $\chi^{2}/\nu$ are
in fact $22/8$ (probability $\sim 5\times10^{-3}$) and $13/10$
(probability $\sim$0.2) for the distribution of fluxes and their
logarithmic values, respectively. The log fluxes follow the normal distribution with normalisation $A = 36.4\pm4.0$, $\sigma = 0.7\pm1.8\times10^{-1}$ and mean, $\mu = 1.1\pm0.2$ with a probability of $\sim 5\times10^{-3}$ of exceeding $22/8$. The best fit values for a normal distribution of the fluxes themselves are $A = 29.1\pm3.4$, $\sigma = 0.8\pm0.1$ and mean, $\mu = 0.5\pm0.1$ with a probability of $\sim 0.2$ of exceeding $13/10$.  Therefore, fluxes statistically prefer a lognormal distribution. To determine more clearly, whether the underlying distribution is indeed log-normal, one would have to simulate light curves with 
both normal and lognormal flux distributions along with the appropriate power spectral density as in \cite{2013MNRAS.433..907E} 
and then perform a full likelihood analysis. This goes beyond the scope of the current proceedings and will be the subject of a 
future paper.

\section{Power Spectral Density}

An important characteristic central to the temporal structure of the variations in the emission of a source is the power spectral 
density (PSD) \cite{2013MNRAS.433..907E, biteau:pastel-00822242}. The PSD represents the density of temporal fluctuations expressed in the frequency domain. For AGNs, the 
PSD is often a power-law ($\propto \nu^{-\alpha}$), with different flaring and quiescent states characterised by different powers. 
\begin{figure}[h!]
  \centering
   \includegraphics[width=0.8\linewidth]{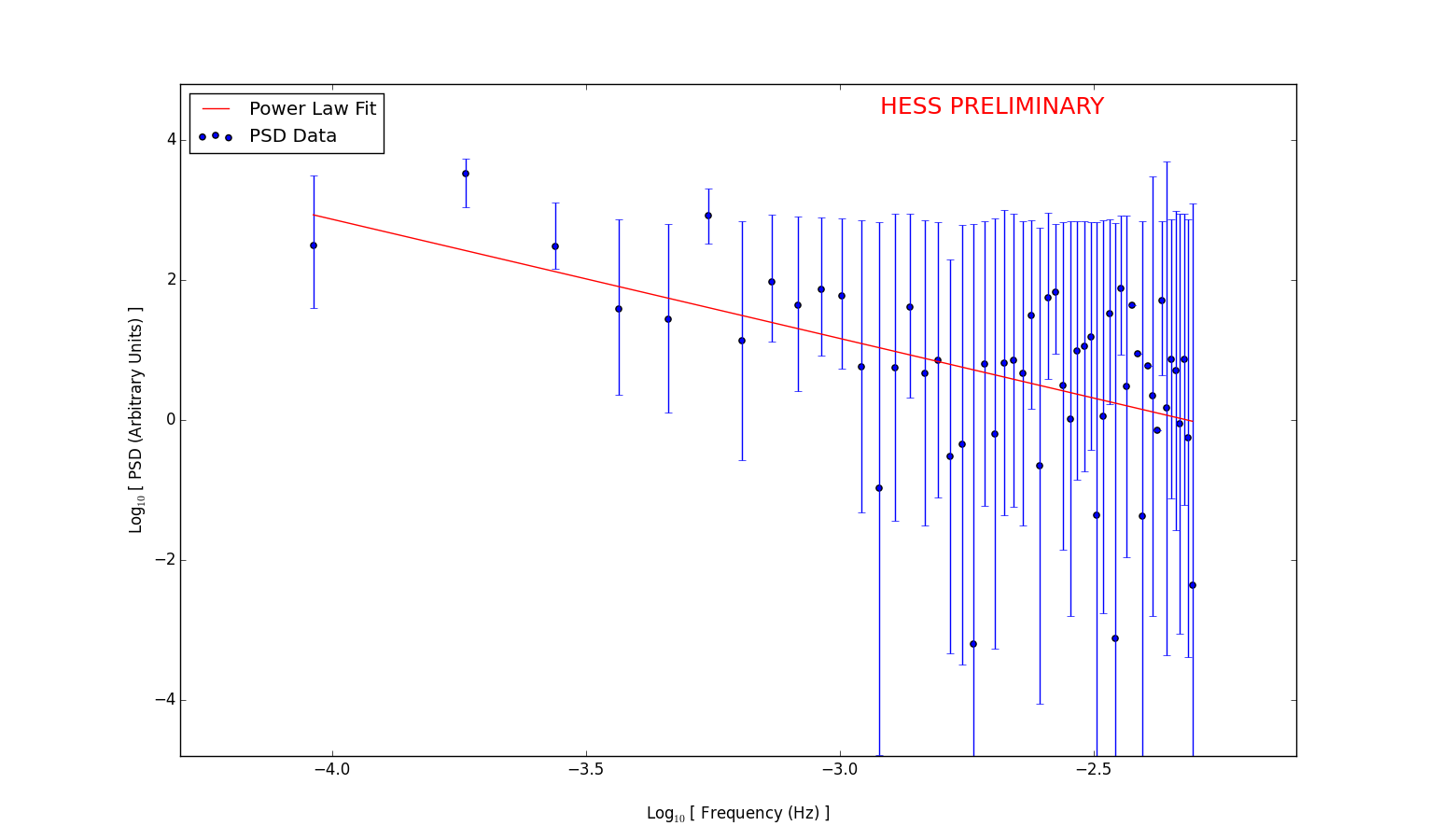}
  \caption{The figure shows the PSD for fluxes $\phi$(> 2 TeV) binned at 4 minutes. The red line represents a single power law fit to the data - this suggests that above $\nu \gtrsim (1-2)\times10^{-3}$ Hz, a simple power law is no longer a good description of the data. The error bars are computed analytically assuming Gaussian uncertainties using \cite{biteau:pastel-00822242}.}
  \label{fig:psdflaremrk5012014}
\end{figure}

In the observations of June 2014, the periodogram is used as an estimator of the PSD \cite{2013MNRAS.433..907E}. The PSD of the data points is computed and can be described by a simple power law given by,
\beqar
P(\nu) = A\ \nu^{-\alpha}
\eeqar,
where A is the proportionality constant and $\alpha$ is the index. The best fit values are $A = 10^{-3.9\pm1.1}$ and $\alpha = -1.7\pm0.4$ with a reduced chisquare value of 1.4 for 52 degrees of freedom. This corresponds to a probability of $\sim 4.5\times10^{-2}$. The best fit is shown by the red line in the figure~\ref{fig:psdflaremrk5012014}. The error bars are computed using the analytical formula in \cite{biteau:pastel-00822242}. This computation assumes that the data have Gaussian uncertainties. The errors are larger for Mrk 501 than for PKS 2155-304 as expected given the comparison of statistics between the two flaring episodes, necessitating simulations for drawing robust conclusions. It is evident from the figure that, such a simple power law red noise spectrum represents a reasonable fit up to 
frequencies of $\approx (1 - 2) \times10^{-3}$ Hz, after which the scatter in the data weakens the constraints. Given the binsize of 4 minutes, it is expected that the fluctuations at frequencies higher than 
$1/2\times(f_{\rm sampling})$ will be suppressed. Furthermore, with the noise in 
the data above this frequency, it is difficult to constrain the shape of the PSD beyond this frequency. The presence of gaps in the lightcurve adds to the difficulty in placing constraints in general without performing simulations. 
The shape depends on the state of emission and thus constrains the mechanisms. In order to do this robustly, one again needs to perform a full likelihood analysis with simulated light curves and test for the likelihood of a simple power law describing the data. 

\section{Conclusions and Discussion}

The flare of June 2014 in Mrk 501 was quite unprecedented in that it showed fast variations (on a few minutes timescales) 
at purely TeV energies. This provides us with the unique opportunity to test the characteristics of the emission processes in 
Mrk~501 at higher energies than seen till date for this source. Our analysis reveals that the TeV flux variations tend to favour 
a lognormal distribution which along with the inferred variability timescales could offer important clues as to a hadronic or 
leptonic origin. In principle, log-normality is suggestive of a multiplicative process and may be a natural outcome of a 
"cascade-like'' hadronic scenario or an intrinsic jet-disk connection, while generally fast variability is more easily accommodated 
in a leptonic context. Combined with observations at other wavelengths such as X-rays, these results will thus help to put 
further constraints on  the magnetic field, the size and extent of the emission region and indeed the relative importance of 
hadronic versus leptonic processes as well as the applicability of different model assumptions, e.g. \cite{2000NewA....5..377A,
2012A&A...541A..31N, 2015ApJ...798....2S}. The inferred power spectral density is consistent with a single power law up to a 
certain frequency beyond which the scatter does not allow the PSD to be well constrained. The index of $\sim\ 1.7$ derived here for Mrk 501 is similar to the red-noise index of $\sim 2$ derived for PKS 2155-304 \cite{2008bves.confE..66S}. The onset of scatter may simply be a result of noise due to background fluctuations or a limitation due to lack of significant bins. Furthermore, presence of gaps in the lightcurve prevent putting definitive constraints.  To assess this in more detail, a more sophisticated statistical treatment 
will be needed which will be presented elsewhere.

\section{Acknowledgments}
The support of the Namibian authorities and of the University of Namibia in facilitating the construction and operation of H.E.S.S. is gratefully acknowledged, as is the support by the German Ministry for Education and Research (BMBF), the Max Planck Society, the German Research Foundation (DFG), the French Ministry for Research, the CNRS-IN2P3 and the Astroparticle Interdisciplinary Programme of the CNRS, the U.K. Science and Technology Facilities Council (STFC), the IPNP of the Charles University, the Czech Science Foundation, the Polish Ministry of Science and Higher Education, the South African Department of Science and Technology and National Research Foundation, and by the University of Namibia. We appreciate the excellent work of the technical support staff in Berlin, Durham, Hamburg, Heidelberg, Palaiseau, Paris, Saclay, and in Namibia in the construction and operation of the equipment. The support of AvH foundation is also acknowledged. This research used lightcurve simulation codes provided by Physics and Astronomy, University of Southampton available at \verb+https://github.com/samconnolly/DELightcurveSimulation+ for computation of the PSD and the analytical expressions from the thesis of Jonathan Biteau for the PSD errors.  
\bibliographystyle{JHEP}
\bibliography{icrcmrk501}

\end{document}